# Multi-Octave Supercontinuum Generation and Frequency Conversion based on Rotational Nonlinearity

**Authors:** John E. Beetar[1†], M. Nrisimhamurty[1†], Tran-Chau Truong[1], Garima C. Nagar[2], Jonathan Nesper[1], Omar Suarez[1], Yi Wu[1,3,4], Bonggu Shim[2] and Michael Chini[1,4,*]

**Affiliations:**

[1]Department of Physics, University of Central Florida, Orlando FL 32816

[2]Department of Physics, Applied Physics and Astronomy, Binghamton University, Binghamton NY 13902

[3]Institute for the Frontier of Attosecond Science and Technology, University of Central Florida, Orlando FL 32816

[4]CREOL, the College of Optics and Photonics, University of Central Florida, Orlando FL 32816

*Correspondence to: Michael.Chini@ucf.edu.

†These authors contributed equally to this work.

**Abstract:** The field of attosecond science was first enabled by nonlinear compression of intense laser pulses to a duration below two optical cycles. Twenty years later, creating such short pulses still requires state-of-the-art few-cycle laser amplifiers to most efficiently exploit "instantaneous" optical nonlinearities in noble gases for spectral broadening and parametric frequency conversion. Here, we show that nonlinear compression can in fact be much more efficient when driven in molecular gases by pulses substantially longer than a few cycles, due to enhanced optical nonlinearity associated with rotational alignment. We use 80-cycle pulses from an industrial-grade laser amplifier to simultaneously drive molecular alignment and supercontinuum generation in a gas-filled capillary, producing more than two octaves of coherent bandwidth and achieving >45-fold compression to a duration of 1.7 cycles. As the enhanced nonlinearity is linked to rotational motion, the dynamics can be exploited for long-wavelength frequency conversion and compressing picosecond lasers.

**Main Text:** The temporal confinement of light to durations close to the optical period (*1*), and the conversion of such few-cycle pulses to extreme ultraviolet and x-ray wavelengths (*2, 3*), has enabled measurement and control of electron dynamics on the sub-femtosecond time scale (*4, 5*). A prerequisite for entry into the single-cycle regime is the coherent generation of octave-spanning spectra, which can be achieved through nonlinear propagation of multi-cycle laser pulses through solid, liquid, or gaseous media. The broadened supercontinuum is a hallmark of self-phase modulation (SPM) (*6*), a nonlinear process governed by the intensity dependent refractive index $\Delta n_i = n_2 I$ associated with the optical Kerr effect.

Key properties of the Kerr nonlinearity have led to the widespread adoption of laser pulse compression schemes based on SPM in gas-filled capillary fibers (*7-10*). As the refractive index changes arise due to field-driven distortions of the valence electronic wave function (*11*), having a natural time scale below 1 fs, the response is effectively instantaneous and the refractive index tracks the time-dependent laser intensity $I(t)$. This in turn imparts a nearly linear spectral chirp onto the pulses during propagation, allowing for compression to nearly bandwidth-limited durations with the use of prism pairs, gratings, or chirped mirrors. Furthermore, the tight binding of valence electrons in inert gases allows the use of relatively high laser intensities before the detrimental onset of ionization-related dispersion and losses.

Generation of the shortest pulses, and broadest supercontinuum spectra, has therefore been traditionally accomplished by driving the SPM process in noble gas-filled capillary fibers using shorter input pulses (*12*). Shorter driving pulses allow exposure of the neutral gas atoms to higher laser intensities, or, alternatively, the use of heavier gases with larger $n_2$, both resulting in larger nonlinear phase shifts $\Delta\phi_{nl}(t) \propto n_2 I(t)$. To this end, significant effort has been invested into the development of few-cycle laser amplifiers (*13, 14*) and two-stage compression schemes (*15-17*), which mitigate the effects of ionization and plasma generation on the efficiency and compressibility of broad supercontinuum spectra.

Here, we suggest a paradigm shift in few-cycle pulse compression, wherein the input pulse duration is chosen to optimize the nonlinearity rather than minimize the ionization. By taking advantage of the much larger rotational contribution to the nonlinear index of refraction of molecules (*18*) and using relatively long laser pulses for which the rotational nonlinearity can be nearly instantaneous, we demonstrate a record >45x pulse compression in a single stage. Moreover, we show that the timescale of the rotational nonlinearity can be used to shift the central frequency of the few-cycle pulses, thereby providing a more-efficient alternative to long-wavelength sources based on parametric amplification (*19, 20*).

For molecular gases, the vibrational and rotational degrees of freedom result in the addition of "delayed" nonlinear responses $\Delta n_d$ caused by the field-induced alignment and stretching of the molecular bonds (*21*). The rotational nonlinearity, which for linear molecules dominates over the vibrational contribution (*22*), has only minimal impact on the propagation of short pulses (below ~100 fs) due to a lack of temporal overlap with the delayed response (*23*). However, it has been exploited to imprint temporal phase shifts onto a second copropagating pulse which arrives during a revival of rotational coherence (*24*). Conversely, long input pulses (100s of fs) experience a prolonged interaction which can coincide with the delayed response duration (*18*), dramatically increasing the magnitude of the induced rotational nonlinearity, as illustrated in Figure 1. For sufficiently long pulses, the enhanced rotational nonlinearity can also be considered to be nearly-instantaneous and can yield a qualitatively similar time-dependent nonlinear response as that of the electronic Kerr nonlinearity. By appropriately choosing the molecular gas according to the input pulse duration, the total nonlinearity can be enhanced more than 10-fold in comparison to an atomic gas with similar nonlinear susceptibility and ionization potential.

To illustrate how the dynamics of the delayed nonlinearity impact the generated supercontinuum, we calculate the spectrum generated from propagation of laser pulses with various input pulse durations (30 fs to 1 ps) in Ar, $N_2$, and $N_2O$ (Figure 2) using a model which incorporates the delayed nonlinearities using a density matrix formalism to obtain the total nonlinear phase shift (*25*). While argon exhibits symmetric spectral broadening throughout, with a spectral bandwidth that quickly shrinks as the input pulse duration increases, both the molecular gases exhibit dramatically different behavior depending on the input pulse duration. Both $N_2$ and $N_2O$ start with a broad supercontinuum at short input pulse durations, which initially shrinks with increasing pulse duration, but is then followed by a strong red-shift and subsequent extension to higher frequencies. The blue-shifted components are initially weak, but gradually increase in strength with further increasing input pulse duration. Notably, the supercontinuum spectra generated in $N_2$ and $N_2O$ with very long pulses remain substantially broader than those generated in Ar due to the enhancement afforded by the large rotational nonlinearity. For the longest pulses considered, molecular alignment approaches the adiabatic condition where $\Delta n_d$ and $\Delta n_i$ have the same qualitative behavior, yielding symmetric spectra with broader bandwidth than their atomic counterparts.

We experimentally demonstrate coherent supercontinuum generation by propagating 400 µJ, 280 fs pulses with a central wavelength of 1025 nm through a 3.5 m long capillary fiber filled with Ar, $N_2$, or $N_2O$ (*26*). Pressure-dependent spectra for Ar and $N_2$ up to 6.2 bar and for $N_2O$ up to 4.4 bar are shown in the Supplementary Material, while the spectra collected at the highest pressures are shown in Figure 3(a-c). The measured spectra exhibit extensions to both the short and long wavelength sides of the input spectrum and amplitude modulations consistent with the SPM process and with numerical propagation simulations (*27*) shown in Figure 3(d-f). In the case of $N_2$ and $N_2O$ the spectral weight is shifted more heavily towards longer wavelengths, resulting in red-shifted central wavelengths of 1094 nm and 1068 nm, respectively. This is unlike typical SPM-broadened spectra, where the generation of frequencies below and above the fundamental frequency occurs at the leading and trailing edges of the pulse, respectively, leading to a symmetric spectrum such as that observed in Ar. Instead, due to the delayed rotational nonlinearity, the zero crossing of the instantaneous frequency shift comes after the pulse's peak (Fig. 1d), leading to the enhanced generation of red-shifted frequencies.

At 6.2 bar, the supercontinuum spectra generated in Ar and $N_2$ cover 200 nm and 485 nm, respectively. These correspond to bandwidth-limited pulse durations of 21 fs (6 cycles at 1026 nm) and 7.4 fs (2 cycles at 1094 nm) for Ar and $N_2$, respectively. In the case of $N_2O$, the supercontinuum generated at 4.4 bar covers two optical octaves and supports a bandwidth-limited pulse duration of 2.5 fs, less than one optical cycle at a central wavelength of 1068 nm. Synthesis of such short pulses requires that the phase accumulated during propagation can be adequately compensated over the entire bandwidth. We separately demonstrate compression of the supercontinuum spectra generated in $N_2$ and $N_2O$ to few-cycle durations in two spectral regions using dispersion-compensating mirrors (*28*) (700-1400 nm) and an acousto-optic programmable dispersive filter (*29*) (1100-1800 nm). The few-cycle pulses characterized by frequency-resolved optical gating (FROG), and their corresponding spectra, are shown in Figure 4. Using dispersive mirrors, we generate 7.2 fs (2 cycles at 1067 nm) pulses in 6.4 bar of $N_2$, and 6.0 fs (1.7 cycles at 1135 nm) pulses in 2.4 bar of $N_2O$, corresponding to a >45-fold compression. With the dispersive filter, we select a region of the spectrum not covered by our dispersive mirrors and demonstrate compression in the long wavelength region of the generated supercontinuum to 10.4 fs (2.3 cycles at 1375 nm) in 3.0 bar of $N_2O$.

The presented results underscore new capabilities associated with the judicious choice of molecular gas media for nonlinear propagation according to the laser pulse parameters. In addition to optimizing the accumulated nonlinear phase for supercontinuum generation, we highlight a mechanism through which the laser's central frequency can be efficiently red-shifted during spectral broadening. Accordingly, for a given molecular gas there exists a range of pulse durations that can provide either symmetric or red-shifted spectral broadening. We therefore identify propagation in molecular gases as an efficient platform for both the compression of long laser pulses, potentially yielding few-cycle pulses directly from industrial-grade picosecond lasers (*30*), and the generation of high-energy, long-wavelength femtosecond pulses in hollow-core fibers (*31*) (see Supplementary Text). Optimistically, the use of larger linear molecules with substantially larger electronic and rotational nonlinearities and longer rotational periods (*32*) could potentially allow nonlinear compression of pulses with durations greater than 10 ps, which can be generated via direct amplification.

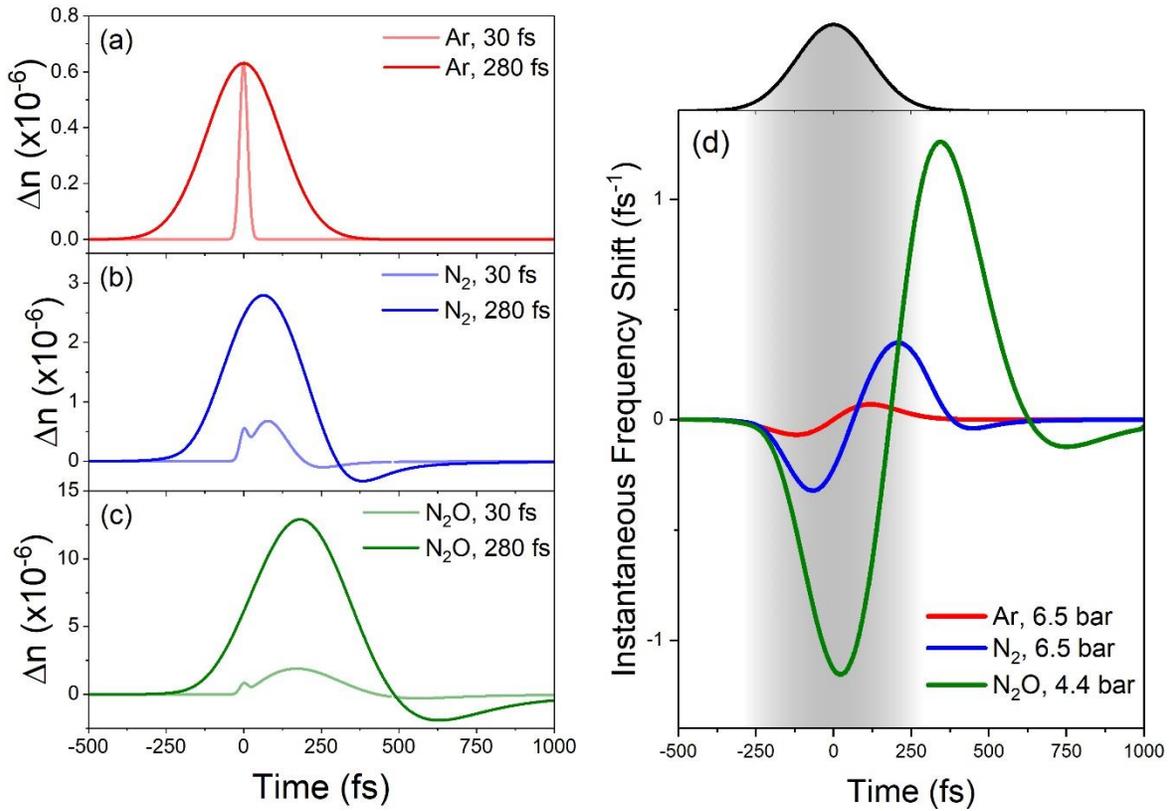

**Figure 1: Nonlinear indices and phase shifts.** The nonlinear indices of refraction of Ar, $N_2$, and $N_2O$ (a-c) are calculated for short (30 fs) and long (280 fs) input pulse durations. Ar and $N_2$ share similar ionization potentials and instantaneous nonlinear refractive indices. Both atomic and molecular systems exhibit an instantaneous response $\Delta n_i$ due to the electronic Kerr nonlinearity, while the delayed rotational response $\Delta n_d$, proportional to the degree of alignment in the molecular ensemble, is present only in the molecular systems. For short input pulses, the effective change in refractive index is approximately the same in Ar and $N_2$ since only the instantaneous contribution takes effect during the period of interaction, whereas the long pulses see a significantly larger $\Delta n$ in $N_2$. The effect is even more pronounced in $N_2O$, which was chosen due to its larger polarizability anisotropy and longer rotational period. The latter plays a crucial role since the degree to which the temporal evolution of the refractive index coincides with the intensity envelope (shown in grey for the 280 fs pulse) determines the shape of temporal phase shift and thus the time dependence of the instantaneous frequency (d).

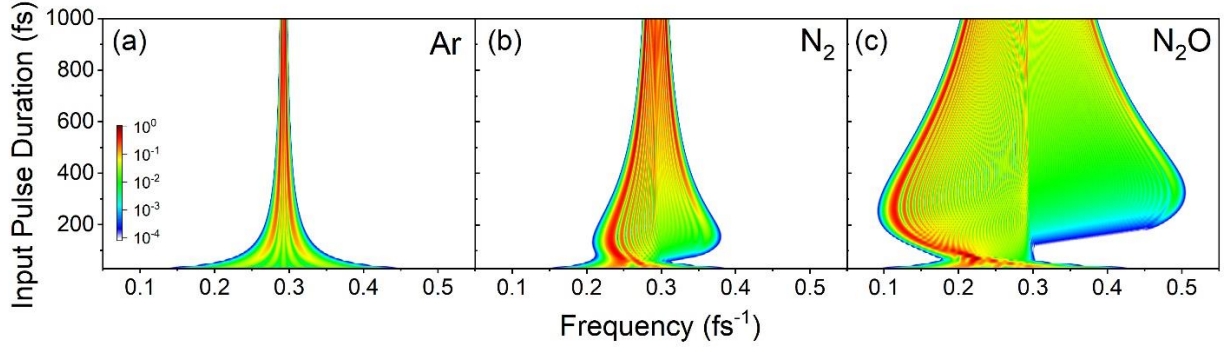

**Figure 2: Pulse duration dependence of spectral broadening.** Comparison of the supercontinuum spectra simulated in (a) Ar (6.5 bar), (b) $N_2$ (6.5 bar) and (c) $N_2O$ (4.4 bar) for different input pulse durations at a fixed intensity of 1 TW/cm$^2$ reveals the dynamics associated with $\Delta n_d$. In the molecular gases, longer pulses lead to larger degrees of molecular alignment and therefore enhanced nonlinearity. At the same time, the molecular alignment is delayed with respect to the pulse. The combination of these two effects leads to purely red-shifted supercontinuum spectra for pulse durations of approximately 100 fs in $N_2$ and 150 fs in $N_2O$. Further increasing the pulse duration shifts the peak of the alignment to coincide with the trailing edge of the pulse, resulting in more symmetric spectra and optimized spectral bandwidth for pulse durations of approximately 150 fs in $N_2$ and 280 fs in $N_2O$. The central frequency of the input laser was 0.29 fs$^{-1}$ in all cases.

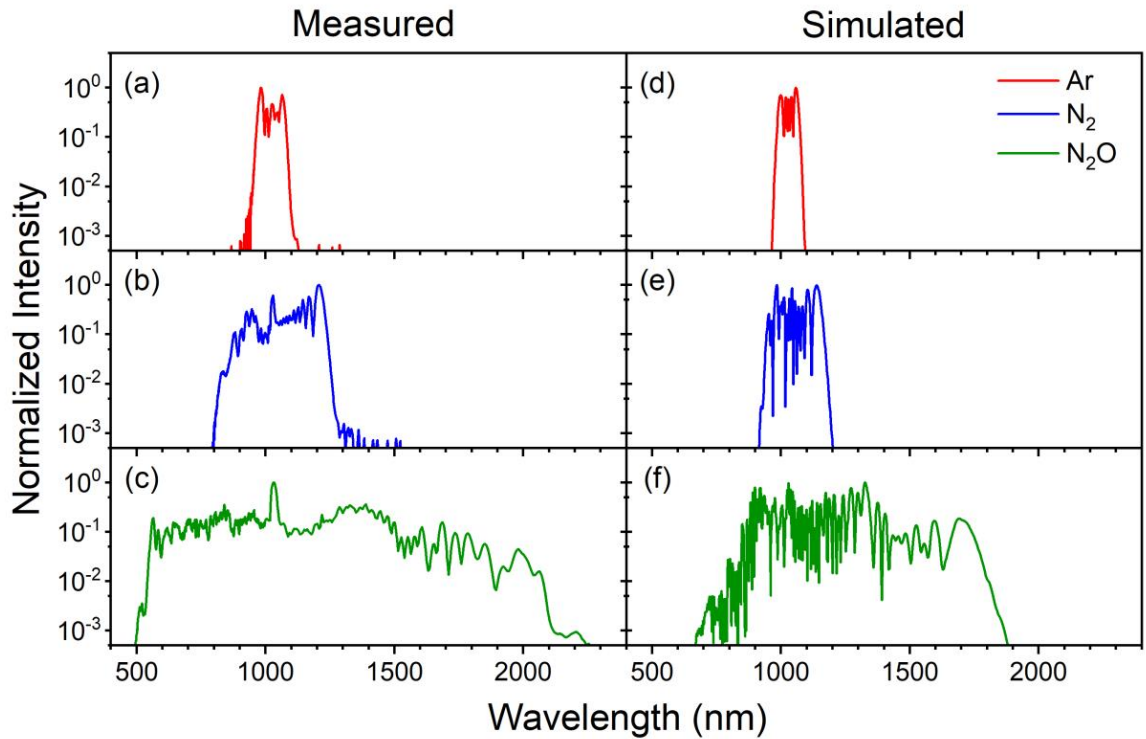

**Figure 3: Supercontinuum spectra.** Supercontinuum spectra measured at the output of the fiber (a-c) show qualitative differences for atomic and molecular gases. Despite similar contributions from electronic Kerr nonlinearity, the spectrum from $N_2$ is 2.4 times broader than that of Ar. Moreover, both the $N_2$ and $N_2O$ spectra display strong well-spaced interference modulations in the longest wavelength components, while the shortest wavelengths are comparatively smooth. This behavior is consistent with the delayed peak of the time-dependent nonlinear refractive index, which occurs on the trailing edge of the pulse as shown in Figure 1. We additionally find qualitative agreement with spectra obtained from numerical propagation simulations (d-f) incorporating both the instantaneous and delayed response of the refractive index.

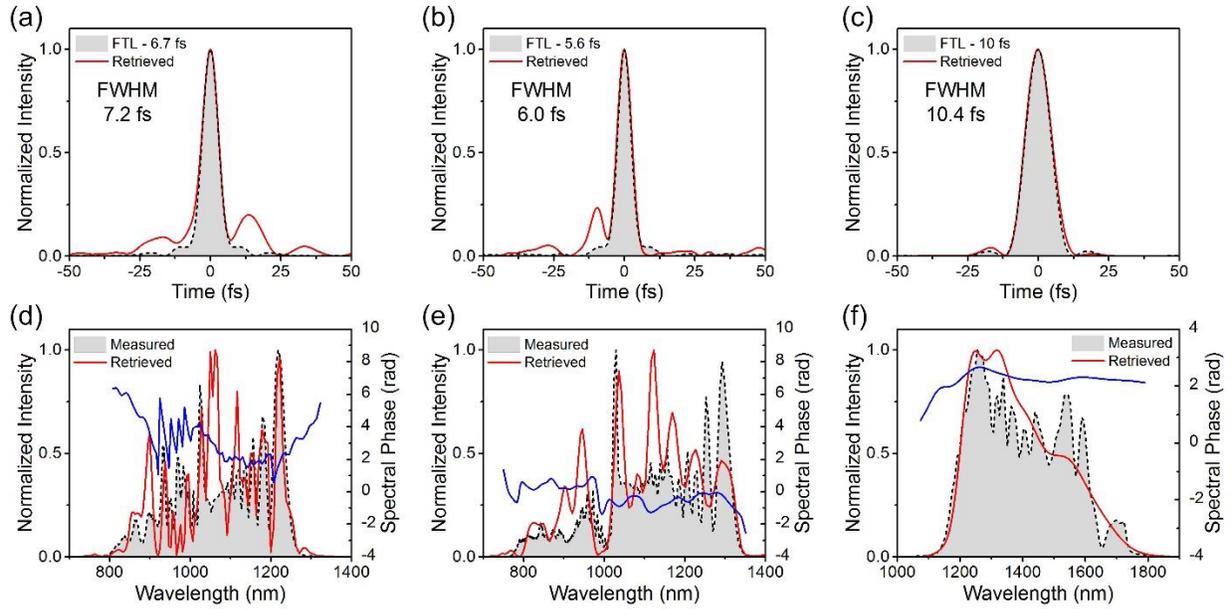

**Figure 4: Pulse compression.** Retrieved temporal (a-c) and spectral intensity (d-f) profiles of few-cycle pulses obtained for the three compression demonstrations: 6.5 bar of $N_2$ using dispersive mirrors (a, d), 2.4 bar of $N_2O$ using dispersive mirrors (b, e), and 3.0 bar of $N_2O$ using a programmable dispersive filter (c, f). The retrieved temporal intensity profiles (red) are shown alongside the bandwidth-limited (FTL) pulse profiles (grey) calculated from the retrieved spectra. The retrieved (red) and measured (grey) spectral intensity profiles are shown along with the retrieved spectral phase (blue). The relatively flat spectral phases demonstrate good dispersion compensation.

**Acknowledgments:** We thank Zenghu Chang for loaning the infrared spectrometer and the acousto-optic programmable dispersive filter for use in the experiments. We also thank Zenghu Chang, Enrique del Barco, and Eric Van Stryland for providing feedback on an earlier version of this manuscript.

**Funding:** This material is based on research supported by the U.S. Department of Energy (DOE), Office of Science, Basic Energy Sciences (BES), under Award No. DE-SC0019291 and by the Air Force Office of Scientific Research (AFOSR) under Award No. FA9550-16-1-0149. B.S. and G.C.N. were supported by the National Science Foundation (NSF) under Award No. PHY-1707237, and the U.S. Air Force Office of Scientific Research (AFOSR) under Award No. FA9550-18-1-0223. Y.W. was supported by the Air Force Office of Scientific Research (AFOSR) under Award No. FA9550-15-1-0037 and by the Defense Advanced Research Projects Agency (DARPA) under Award No. D18AC00011.

**Author contributions:** J.E.B., M.N., and M.C. conceived and designed the study. J.E.B., M.N., C.T., J.N., and O.S. performed the experiments and analyzed experimental data under the supervision of M.C. Y.W. provided guidance for pulse compression and characterization using the acousto-optic programmable dispersive filter. J.E.B. and M.C. performed the model calculations. G.C.N. performed the numerical simulations under the supervision of B.S. J.E.B., M.N., B.S., and M.C. wrote the manuscript, with assistance from all authors.

# Supplemental Material for:
# Multi-Octave Supercontinuum Generation and Frequency Conversion based on Rotational Nonlinearity

**Authors:** John E. Beetar[1†], M. Nrisimhamurty[1†], Tran-Chau Truong[1], Garima C. Nagar[2], Jonathan Nesper[1], Omar Suarez[1], Yi Wu[1,3,4], Bonggu Shim[2] and Michael Chini[1,4,*]

Experimental Setup
The experiments were performed by focusing the output of a Yb: KGW laser amplifier (Light Conversion PHAROS – 1025 nm, 280 fs) with maximum pulse energy and average power of 400 µJ and 20 W, respectively, onto the entrance of a stretched hollow-core capillary fiber (few-cycle, Inc. – 3.5 m length, 500 µm inner diameter) filled with Ar, $N_2$ and $N_2O$. The generated supercontinuum spectra were characterized at the output of the fiber using a set of fiber-coupled spectrometers, described in more detail below. For pulse compression, spectral dispersion was compensated with either a combination of chirped mirrors (Ultrafast Innovations PC1632) and $CaF_2$ glass or an acousto-optic programmable dispersive filter (AOPDF, Fastlite Dazzler UHR 900-1700). Temporal characterization of the few-cycle pulses was performed with a home-built second harmonic generation frequency resolved optical gating (SHG FROG) device in single-shot geometry.

Spectral Characterization
Due to the poor response of silicon photodetectors for wavelengths above 1 µm, we used multiple spectrometers to capture the full spectral bandwidth. For all measurements in Ar and $N_2$, the spectra were collected using both a Si-based VIS-NIR spectrometer (Ocean Optics HR 2000+ES) spanning 180-1100 nm and InGaAs-based spectrometer (Ocean Optics FLAME NIR) spanning 940-1660 nm. For the broader spectra generated from $N_2O$, we used an additional cooled InGaAs-based spectrometer (Spectral Evolution IF2500) spanning 1060-2600 nm. Using detector response curves provided by the manufacturers we corrected the collected spectra, interpolated the data to a common wavelength axis, and stitched them in a shared region where both spectrometers are reasonably efficient. Corrected and stitched spectra collected in Ar, $N_2$, and $N_2O$ for various gas pressures are shown in Figure S1. Individual spectra collected using all three spectrometers for 4.4 bar of $N_2O$, along with the detector response curves, are shown in Figure S2 to indicate how the spectra were stitched.

Temporal Characterization
We characterized few-cycle pulses compressed from the generated supercontinuum spectra under three different conditions. The measured and reconstructed FROG traces are shown, along with the retrieved spectra and temporal intensity profiles, in Figures S3-S5. The specific laser and gas parameters used in each case are given in Table S1. In $N_2$ at a pressure of 6.5 bar, we obtain the best dispersion compensation when using 8 pairs of chirped mirrors (providing a group delay dispersion of approximately -90 $fs^2$ per pair) and 20 mm of $CaF_2$ glass (with a group velocity dispersion of approximately 19 $fs^2$/mm at 1025 nm), resulting in pulses with a FWHM duration of 7.2 fs (2 cycles at the central wavelength of 1067 nm). In $N_2O$ at a pressure of 2.4 bar, we compress comparatively lower-energy pulses using 5 pairs of chirped mirrors and 1.5 mm of $CaF_2$, obtaining pulses with a FWHM duration of 6.0 fs (1.7 cycles at 1135 nm). We also compress a portion of the $N_2O$ spectrum not covered by our chirped mirrors (obtained using the full pulse energy at 3.0 bar) using the AOPDF, after which we obtain pulses with a FWHM duration of 10.4 fs (2.3 cycles at 1375 nm). Due to the low damage threshold of the nonlinear crystal inside the AOPDF, only a small fraction (5 µJ) of the total pulse energy was compressed using the AOPDF.

Delayed Nonlinear Response Model
We model the delayed nonlinear response using a density matrix formalism introduced by Y. -H. Chen et al. (25) and incorporate it with the electronic Kerr nonlinearity to calculate the induced nonlinear phase shift. To capture the effect of the field-induced molecular alignment of $N_2$ and $N_2O$ on the refractive index, the time-dependent ensemble-averaged molecular polarizability $\langle \alpha \rangle_t$ is defined for directions along and perpendicular to the incident electric field. Here, $\langle \alpha \rangle_t = \langle \hat{e} \cdot \bar{\bar{\alpha}} \cdot \hat{e} \rangle$, where $\hat{e}$ is a unit vector along the laser polarization direction, and $\bar{\bar{\alpha}}$ is the polarizability tensor. For linear molecules with symmetry about the z-axis, the only nonzero components of $\bar{\bar{\alpha}}$ are $\alpha_{zz} = \alpha_{\parallel}$, $\alpha_{xx} = \alpha_{yy} = \alpha_{\perp}$ and the change in refractive index due to molecular orientation can be written as:

$$\Delta n_d(t) = \frac{2\pi N}{n_0} \Delta \alpha \left( \langle \cos^2 \theta \rangle_t - \frac{1}{3} \right) \tag{S1}$$

where $N$ is the number density of molecules, $\Delta \alpha = \alpha_{\parallel} - \alpha_{\perp}$, and $\cos^2 \theta$ describes the degree of alignment of a particular molecule's axis with respect to the laser polarization direction. The ensemble-averaged alignment is defined as $\langle \cos^2 \theta \rangle_t = Tr(\boldsymbol{\rho}(t) \otimes \cos^2 \theta) = \rho_{kl} \langle l | \cos^2 \theta | k \rangle$, where $\boldsymbol{\rho}(t)$ is the density matrix, and $|k\rangle = |j, m\rangle$ and $|l\rangle = |j', m'\rangle$ are the rotational eigenstates of the field-free Hamiltonian, described by the quantum numbers $j$ and $m$ corresponding to the total rotational angular momentum and the component of angular momentum along the polarization direction, respectively. Assuming that the molecules occupy the ground electronic and vibrational states at room temperature, the first-order perturbation to the density matrix can be calculated as:

$$\left[\boldsymbol{\rho}^{(1)}(t)\right]_{kl} = -\frac{i}{\hbar} \int_{-\infty}^{t} d\tau [\hbar(\tau), \boldsymbol{\rho}^{(0)}]_{kl} e^{-(i\omega_{kl}+\gamma_{kl})(t-\tau)} \tag{S2}$$

where $\hbar = -\mathbf{p} \cdot \mathbf{E}$ is the perturbation Hamiltonian, $\mathbf{p} = \bar{\bar{\alpha}} \cdot \mathbf{E}$ is the induced dipole moment, and $\mathbf{E}(t) = \varepsilon(t) \cos(\omega t)$ is the time-dependent laser electric field with envelope $\varepsilon(t)$ and central frequency $\omega$. The resonance frequencies $\omega_{kl} = (E_k - E_l)/\hbar$ and dephasing rates $\gamma_{kl}$ correspond to transitions between rotational states $|k\rangle$ and $|l\rangle$ with energies $E_k = hcBj(j+1)$ and $E_l = hcBj'(j'+1)$, where $B$ is the rotational constant. $\boldsymbol{\rho}^{(0)}$ is the zeroth order density matrix describing distribution of rotational states for $t \to -\infty$.

The commutator matrix element in Eq. (S2) is $[\hbar(\tau), \boldsymbol{\rho}^{(0)}]_{kl} = \left(\boldsymbol{\rho}_l^{(0)} - \boldsymbol{\rho}_k^{(0)}\right)\hbar_{kl}$, and the perturbation Hamiltonian is given by $\hbar_{kl} = -\Delta\alpha|\mathbf{E}|^2 \langle k| \cos^2 \theta |l\rangle - \alpha_{\perp} \delta_{kl} |\mathbf{E}|^2$, where $\delta_{kl}$ is the identity matrix. Since the rotational eigenstates of a rigid rotor are spherical harmonics, the matrix elements $\langle k| \cos^2 \theta |l\rangle$ are nonzero only when $m' = m$ and $j' = j$ or $j' = j \pm 2$. The commutator matrix element also vanish for diagonal elements, and Eq. (S2) can be rewritten as:

$$\rho_{j,j-2,m}^{(1)}(t) = -\frac{i}{2\hbar}\left(\rho_{j,m}^{(0)} - \rho_{j-2,m}^{(0)}\right)\Delta\alpha Q_{j,j-2} e^{-(i\omega_{j,j-2}+\gamma_{j,j-2})t} \int_{-\infty}^{t} d\tau \varepsilon^2(\tau) e^{(i\omega_{j,j-2}+\gamma_{j,j-2})\tau} \tag{S3}$$

where $Q_{j,j'} = \langle j, m| \cos^2 \theta |j', m'\rangle$ and $\rho_{j,m}^{(0)} = D_j e^{-hcBj(j+1)/k_B T}/Z$, with rotational partition function $Z$ and statistical weighting factors $D_j$ determined by the nuclear spin statistics. Noting that $\rho_{j,j+2,m}^{(1)}$ is the complex conjugate of $\rho_{j,j-2,m}^{(1)}$, the ensemble-averaged alignment can then be calculated from the density matrix as:

$$\langle \cos^2 \theta \rangle_t = \frac{1}{3} - \frac{2}{15\hbar} \frac{j(j-1)}{2j-1} \left( \rho_j^{(0)} - \rho_{j-2}^{(0)} \right) \Delta\alpha \, \text{Im} \left( e^{(i\omega_{j,j-2}+\gamma_{j,j-2})t} \int_{-\infty}^{t} d\tau \varepsilon^2(\tau) e^{(-i\omega_{j,j-2}+\gamma_{j,j-2})\tau} \right) \tag{S4}$$

To obtain an analytical form of Eq. (S4), we consider a pulse envelope of the form $\varepsilon(t) = E_0 \cos\left(\frac{\pi}{2}\frac{t}{\tau_0}\right)$ for $|t| < \tau_0$ and $\varepsilon(t) = 0$ for $|t| > \tau_0$, and integrate Eq. (S4). Typical values of $\langle \cos^2 \theta \rangle_t$ calculated under our experimental conditions deviate from the isotropic ensemble average $\langle \cos^2 \theta \rangle_t = \frac{1}{3}$ on the order of $10^{-2}$, justifying the use of lowest-order perturbation theory.

Equation (S1) neglects the instantaneous distortions of the electronic wave function, which give rise to the electronic Kerr nonlinearity. We therefore include the instantaneous Kerr term $\Delta n_i = n_2 I(t)$ along with the delayed response given by Eq (S1) to obtain the total nonlinear phase shift associated with an effective propagation length $L_{eff}$:

$$\Delta\phi = \frac{2\pi L_{eff}}{\lambda}(\Delta n_i + \Delta n_d) \tag{S5}$$

where $\lambda$ is the input laser central wavelength. The electronic Kerr nonlinear coefficients $n_2$ for Ar, $N_2$, and $N_2O$, as well as the values for the polarizability anisotropies $\Delta\alpha$ and the rotational constants $B$, were taken from Ref. (*18*). As in Ref. (*25*), we found it necessary to increase the value of $\Delta\alpha$ by a factor of 2 in order to obtain agreement with experimental results.

The plots in Figs. 1 and 2 of the main text were generated based on the nonlinear phase shift given in Eq. (S5). Detailed simulations including linear and nonlinear propagation in the capillary are described in detail below.

Capillary Propagation Simulations
For the comparisons with experimental spectra in Fig. 3 of the main text, we numerically propagate the laser electric fields in the gas-filled capillary using the nonlinear envelope equation (NEE) with a modal expansion for the radial coordinates (*3, 27, 33-35*). We model the linearly polarized beam with cylindrical symmetry around the z-axis in the moving pulse frame, written as $E(r,t,z) = \frac{1}{2}\mathcal{E}(r,t,z)\exp(-i\omega_0 t) + \text{c.c.}$, where $\mathcal{E}(r,t,z) = \sum_n \mathcal{E}_n(t,z) J_0(k_{\perp,n}r)$ is the complex field envelope, $\mathcal{E}_n(t,z)$ is the complex modal amplitude, and $J_0(k_{\perp,n}r)$ is the normalized zeroth-order Bessel function (*34*). The number of Bessel modes used in the simulations is 19. The wave number is $k_{\perp,n} = u_n/a$, where $u_n$ is $n^{\text{th}}$ root of the equation $J_0(u_n) = 0$ and $a$ is the radius of capillary, 250 μm (*27*). First, each modal amplitude is propagated considering dispersion and loss in the frequency domain. Second, after reconstructing the electric field by superposing the Bessel components, nonlinear propagation is solved (also in the frequency domain) considering nonlinear Kerr effect, nonlinear absorption, and plasma defocusing and absorption, which result in coupling of different Bessel modes.

The instantaneous and delayed nonlinearities are modeled according to the expression (*21*):

$$\Delta n = n_2(1-\xi)I(t) + n_2\xi \int_{-\infty}^{t} d\tau\, \mathcal{R}(t-\tau)I(\tau). \tag{S6}$$

The first term in Eq. S6 is the instantaneous electronic response and the second term is the delayed response with fraction of $\xi$. The function $\mathcal{R}(t) = \mathcal{R}_0 \exp(-\Gamma t) \sin(\omega_R t)$ models the molecular rotational response with the characteristic time $\Gamma^{-1}$ and frequency $\omega_R$, where $\mathcal{R}_0 = (\Gamma^2 + \omega_R^2)/\omega_R$. The values of the parameters $\xi, \Gamma$, and $\omega_R$ were determined from fits of $\mathcal{R}(t)$ to the calculated delayed response given in Eq. (S1).

The propagation equation is coupled with the plasma equation:

$$\frac{\partial \rho}{\partial t} = W_{mpi}(\rho_0 - \rho) + W_{ci}\rho - \frac{\rho}{\tau_r}, \tag{S7}$$

where $\rho$ is the free electron density and $\rho_0$ is the density of neutral molecules. Here, the first term on the right hand side describes plasma generation due to multiphoton ionization, the second term describes collisional ionization (*36*) and the third term accounts for the plasma recombination with the electron recombination time $\tau_r$.

In all simulations, the input beam is assumed to be a spatio-temporal Gaussian and its spot size radius is set to be 0.645 times the capillary radius for the maximum coupling efficiency. We perform calculations on a radial grid with step size of 1 μm, and in a time window of 2.5 ps with 4096 points, for all the simulations except for the 800 fs pulse which was performed with an extended time window of 6 ps. Figure S6 shows example plots of the propagation-dependent peak intensity and on-axis (r = 0) spectrum obtained from the capillary simulations. The peak intensity plot shows intensity oscillations due to interference of different modes, while the spectrum plot shows large broadening mainly due to self-phase modulation. Fig. S7 shows the calculated normalized on-axis spectra for $N_2$ (red line) and $N_2O$ (blue line) at the capillary output (350 cm), along with the input laser spectrum (black line). These calculations were performed at pressure of 6.5 bars for $N_2$ and 4.4 bars for $N_2O$. Due to the lack of available data on the dispersion for $N_2O$, we use the dispersion parameters of $N_2$ for the $N_2O$ calculation. Fig. S8 compares the calculated normalized spectra for $N_2O$ with dispersion of $N_2$ (blue line) and without dispersion (red line).

To demonstrate the potential of molecular gases for pulse compression and frequency conversion, we performed additional simulations using pulses having longer duration and longer wavelength. We have targeted pulse parameters which are consistent with commercially available Yb:YAG thin disk (*30*) and Tm:fiber (*37*) laser amplifiers. Specifically, we numerically propagate 800 fs pulses with a central wavelength of 1030 nm in 4.4 bar of $N_2O$, and 200 fs pulses with a central wavelength of 1.95 μm in 6.5 bar of $N_2O$, using the same capillary length and inner core diameter as in the experiments, and using the dispersion parameters of $N_2$.

The spectra obtained at the capillary output from propagating the 800 fs input pulses (Fig. S9) exhibits significant broadening which reduces the bandwidth-limited pulse duration to 11.5 fs (3.2 cycles at 1080 nm). The temporal profile obtained after propagation, shown in Fig. S10a, shows a degree of self-compression manifested in a narrow 16 fs peak with approximately 5 percent of the peak power of the transform-limited pulse. Obtaining few-cycle pulses requires that the spectral phase (blue line in Fig. S9) can be effectively compensated. By adding a quadratic phase function to compensate only the group delay dispersion, a relatively clean pulse with duration of 24 fs can be obtained (Fig. S10b). Higher-order phase compensation, for example using pulse shaper technology, could enable the ~70-fold compression close to the bandwidth limit. In comparison to existing techniques for compression picosecond pulses (*38*), spectral broadening in molecular gas-filled capillaries has significant advantages in terms of compression factor and efficiency.

Propagation simulations using 200 fs input pulses with 1.95 μm central wavelength (Fig. S11), on the other hand, exhibit a substantially red-shifted spectrum with a new central wavelength of 2.46 μm. The bandwidth-limited pulse duration is reduced to 11.2 fs, corresponding to 1.36 optical cycles. In comparison to parametric sources, the frequency conversion is highly efficient, with more than 92% of the pulse energy contained in the red-shifted part of the spectrum with wavelength longer than 2.0 μm. We therefore suggest that spectral broadening in molecular gas-filled capillaries could be a simple and highly-efficient alternative to few-cycle, long-wavelength sources based on optical parametric amplification (*19, 20*).

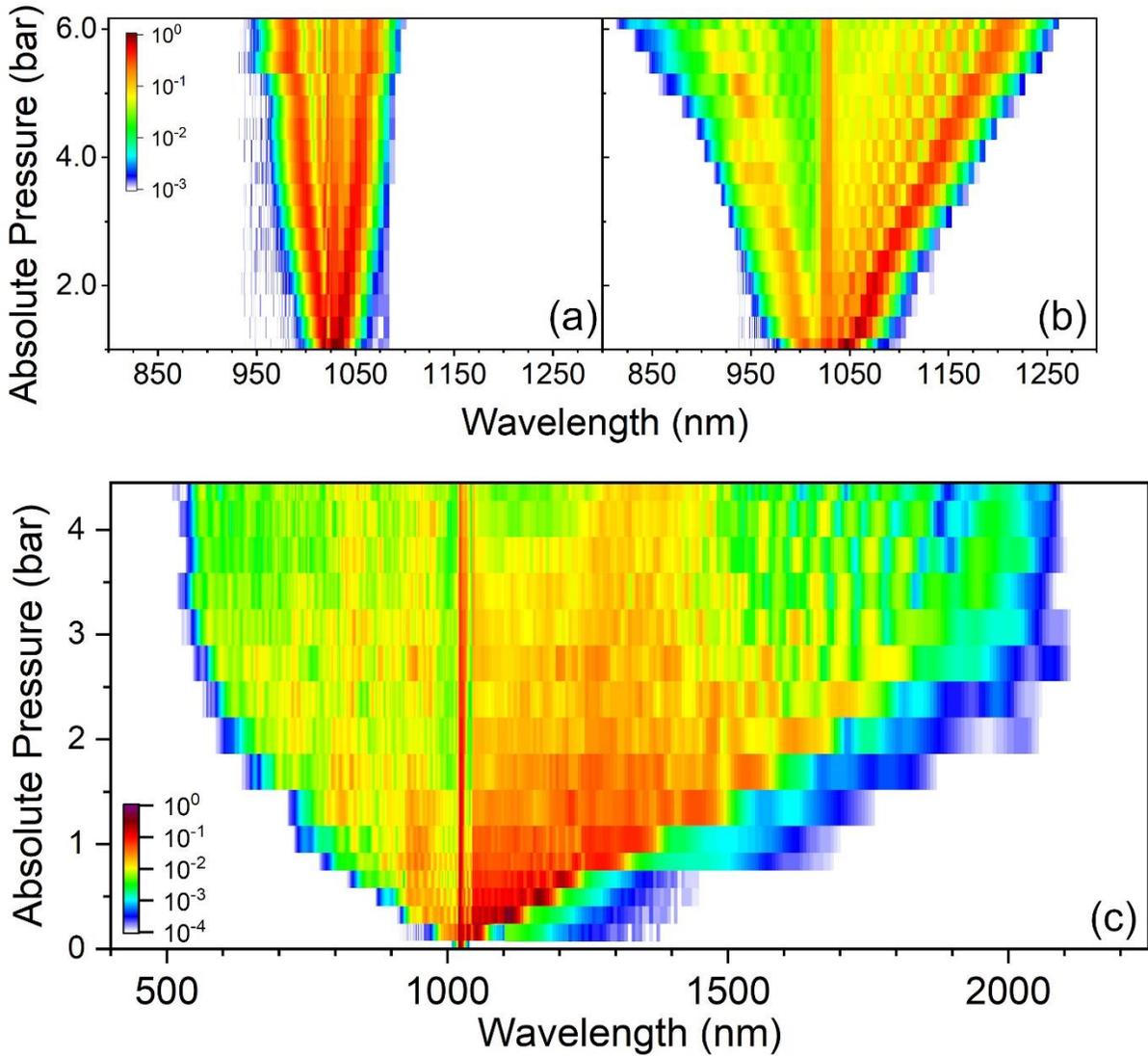

**Fig. S1.** Pressure-dependent spectra measured in (a) Ar, (b) $N_2$, and (c) $N_2O$ using 400 µJ, 280 fs input pulses. By comparing the supercontinuum spectra generated in Ar and $N_2$, which have similar electronic Kerr nonlinearity coefficients $n_2$, one can clearly see the impact of the rotational nonlinearity. The broadest spectra generated in $N_2O$ cover two octaves of spectral bandwidth.

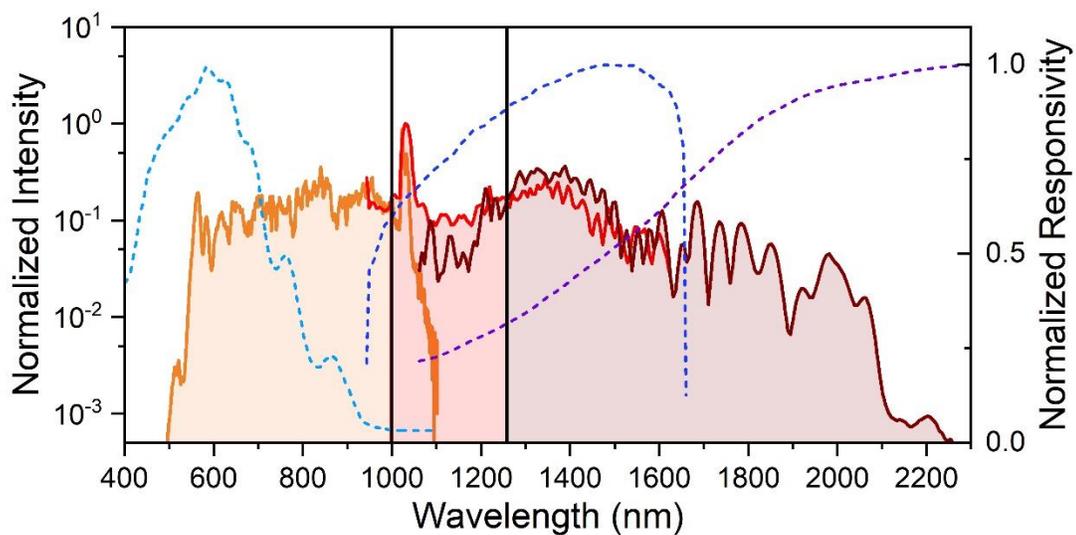

**Fig. S2.** Detector efficiency-corrected spectra measured in $N_2O$ using three different spectrometers. The black solid lines show the wavelengths at which the individual spectra were stitched together. The dotted lines show the normalized responsivity for each spectrometer.

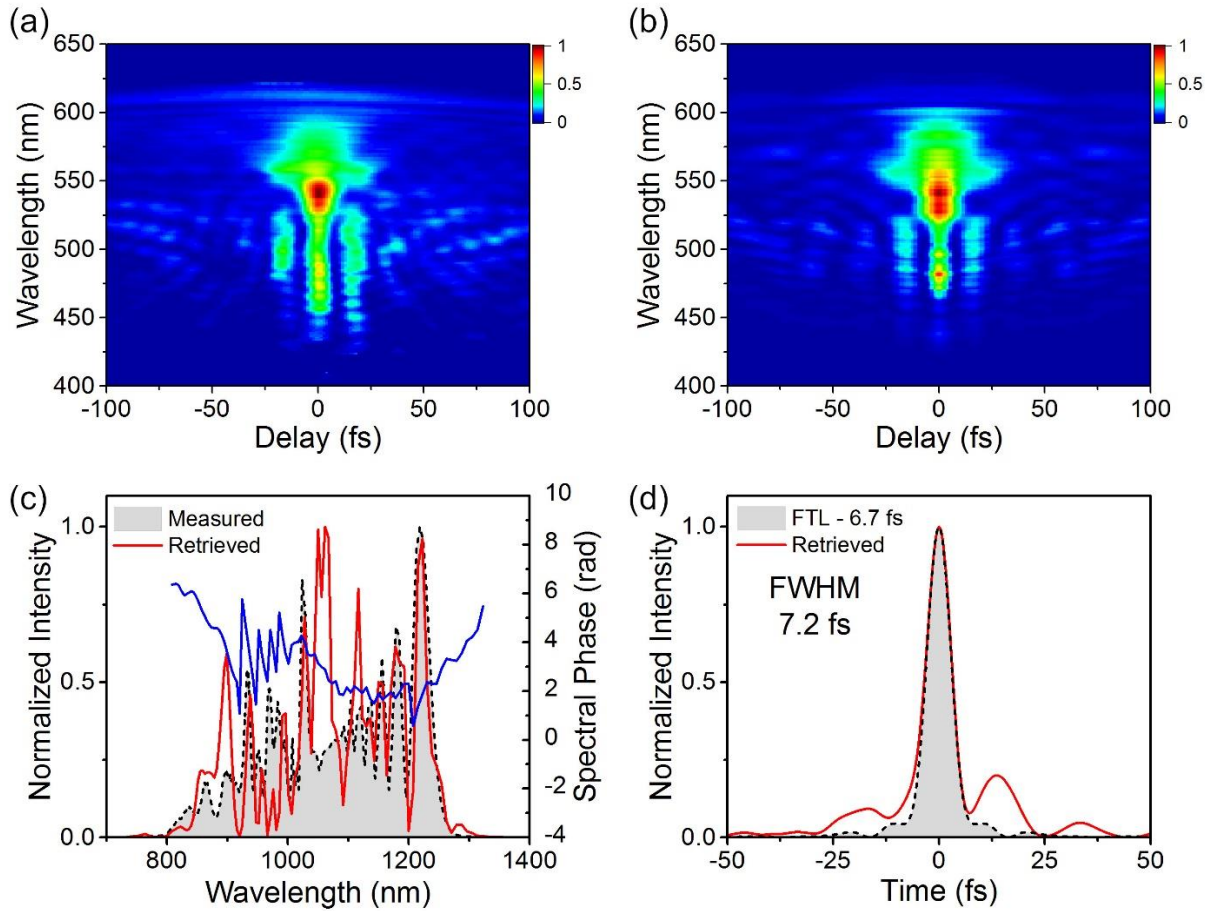

**Fig. S3.** FROG reconstruction of pulses compressed from the $N_2$ supercontinuum using chirped mirrors. (a, b) Measured and reconstructed FROG traces. (c) Measured (grey) and retrieved (red) spectra. The retrieved spectral phase is shown in blue. (d) Bandwidth-limited (grey) and retrieved (red) intensity profiles.

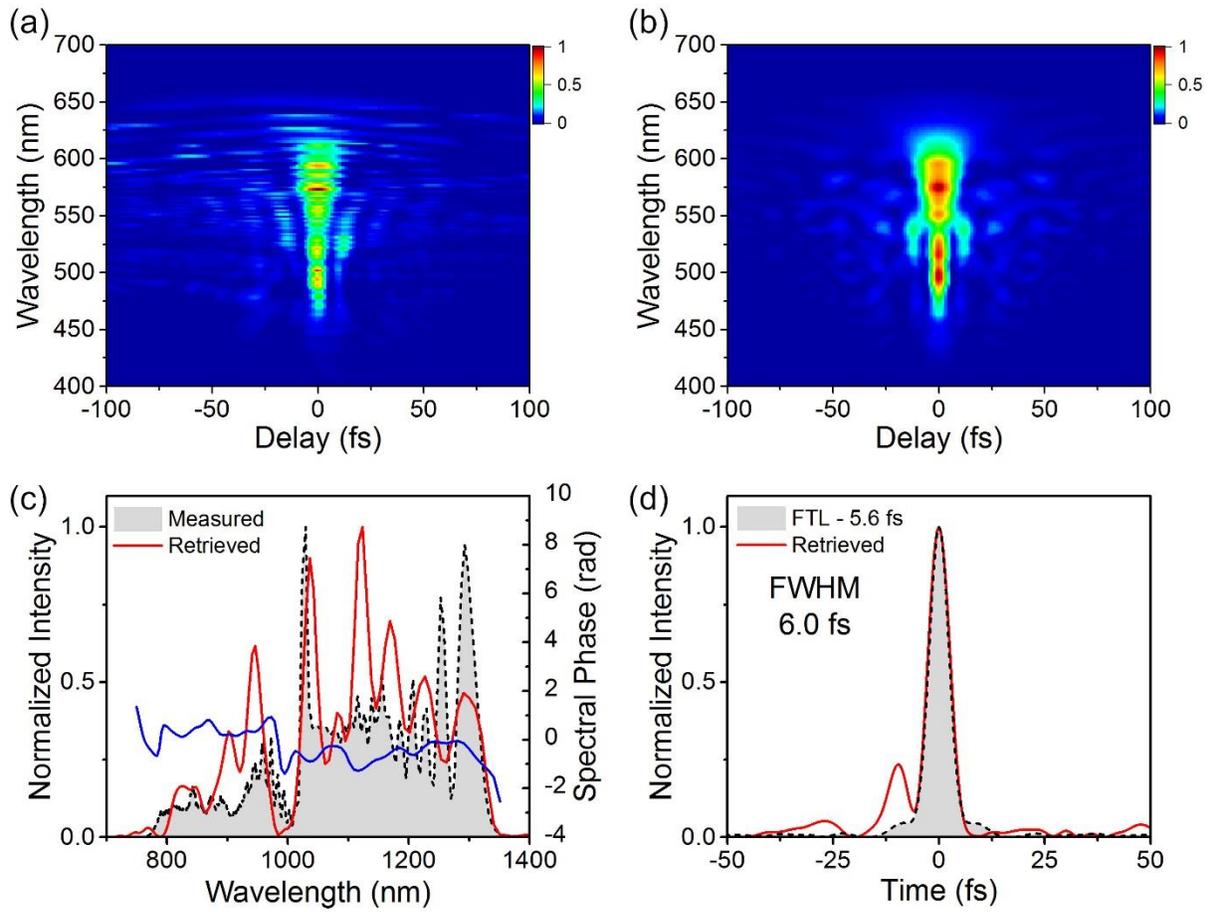

**Fig. S4.** FROG reconstruction of pulses compressed from the N$_2$O supercontinuum using chirped mirrors. (a, b) Measured and reconstructed FROG traces. (c) Measured (grey) and retrieved (red) spectra. The retrieved spectral phase is shown in blue. (d) Bandwidth-limited (grey) and retrieved (red) intensity profiles.

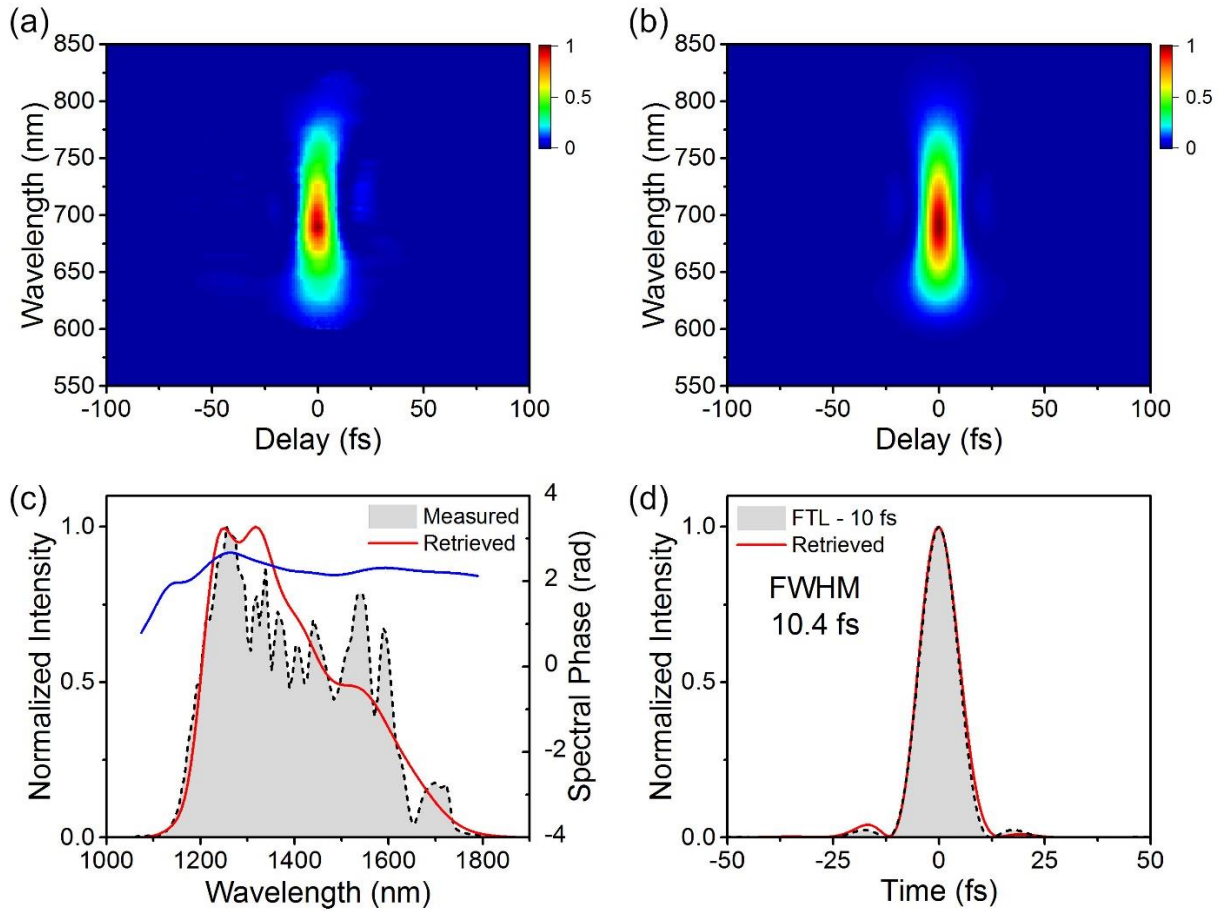

**Fig. S5.** FROG reconstruction of pulses compressed from the long-wavelength portion of the N$_2$O supercontinuum using an AOPDF. (a, b) Measured and reconstructed FROG traces. (c) Measured (grey) and retrieved (red) spectra. The retrieved spectral phase is shown in blue. (d) Bandwidth-limited (grey) and retrieved (red) intensity profiles.

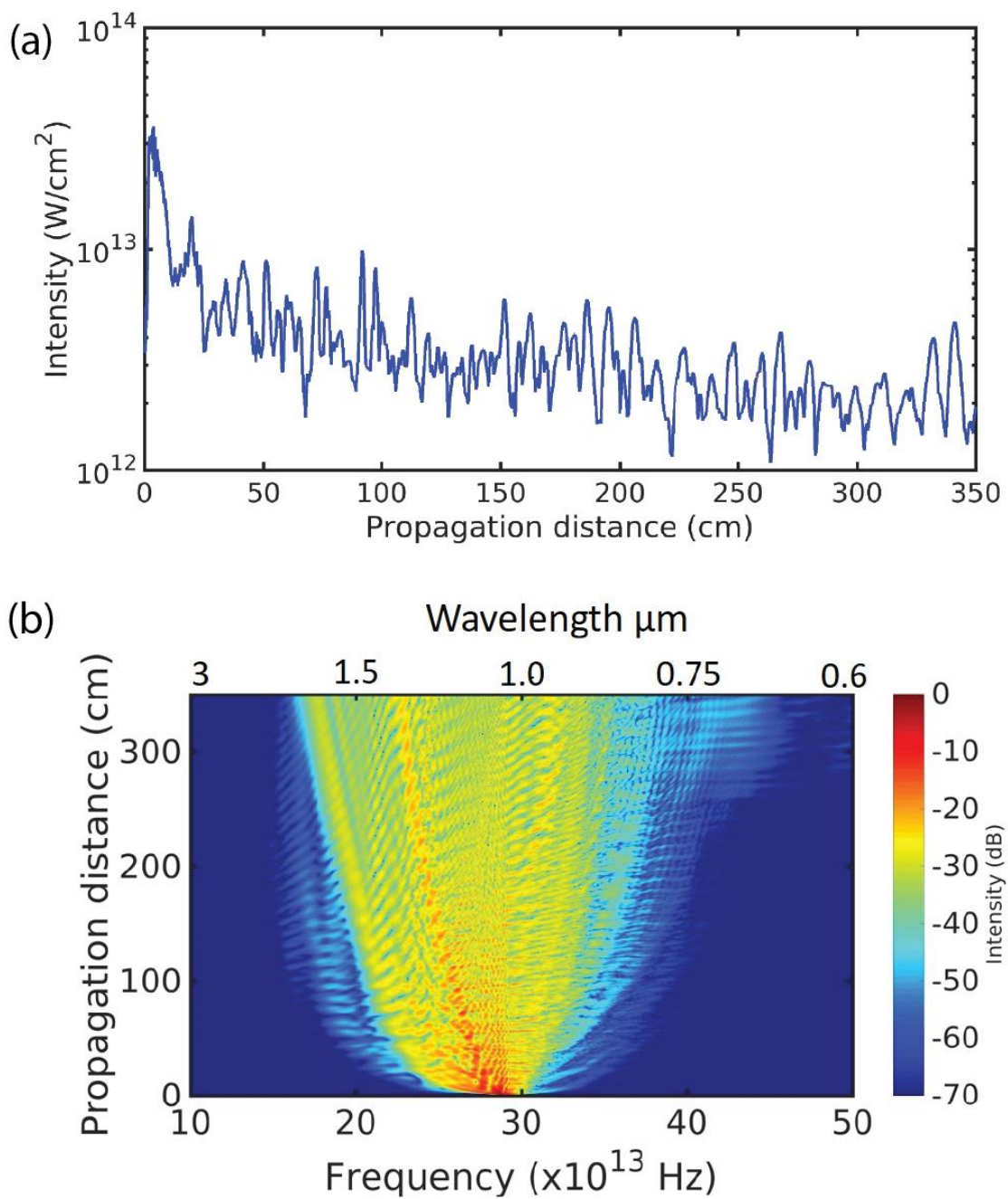

**Fig. S6.** Examples of calculated (a) peak intensity vs. distance and (b) on-axis (r = 0) spectra vs. distance for $N_2O$-filled capillary.

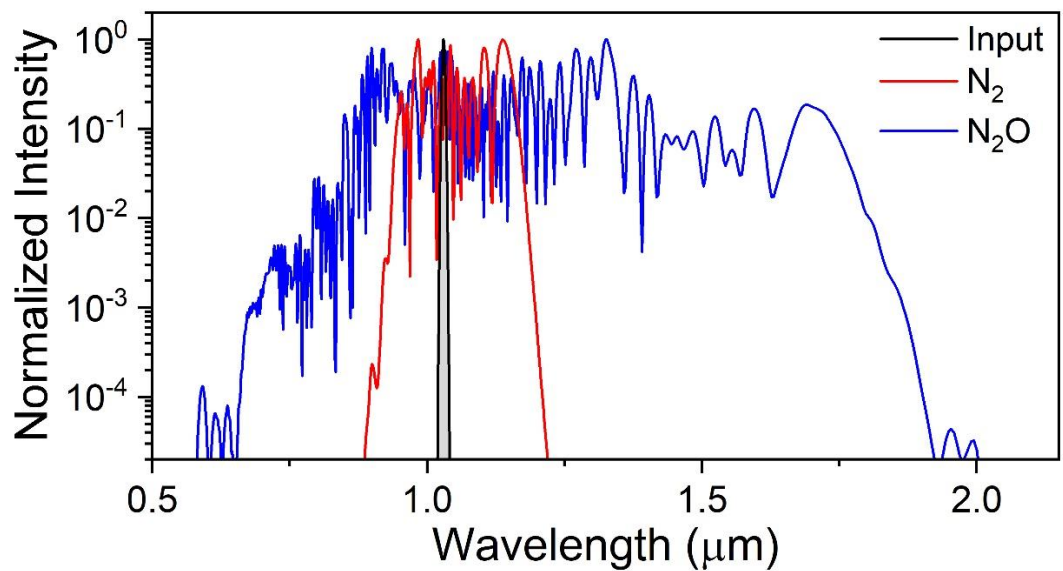

**Fig. S7.** Normalized calculated spectra after propagation in $N_2$ (red line) with 6.5 bars pressure and $N_2O$ (blue line) with 4.4 bars pressure. The black line shows the input laser spectrum.

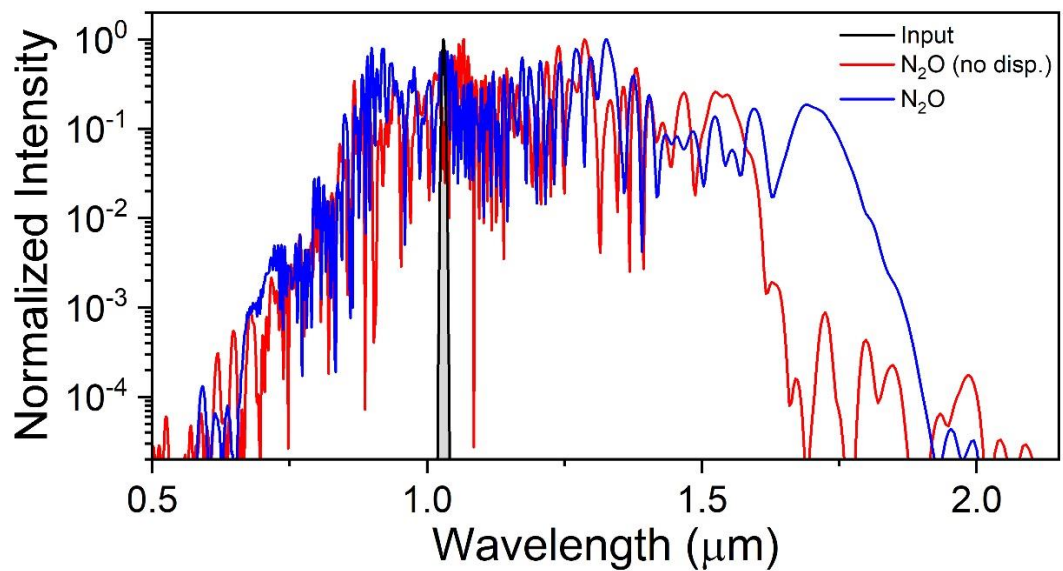

**Fig. S8.** Normalized calculated spectra after propagation in $N_2O$ with dispersion of $N_2$ (blue line) and without dispersion (red line). The black line shows the input laser spectrum.

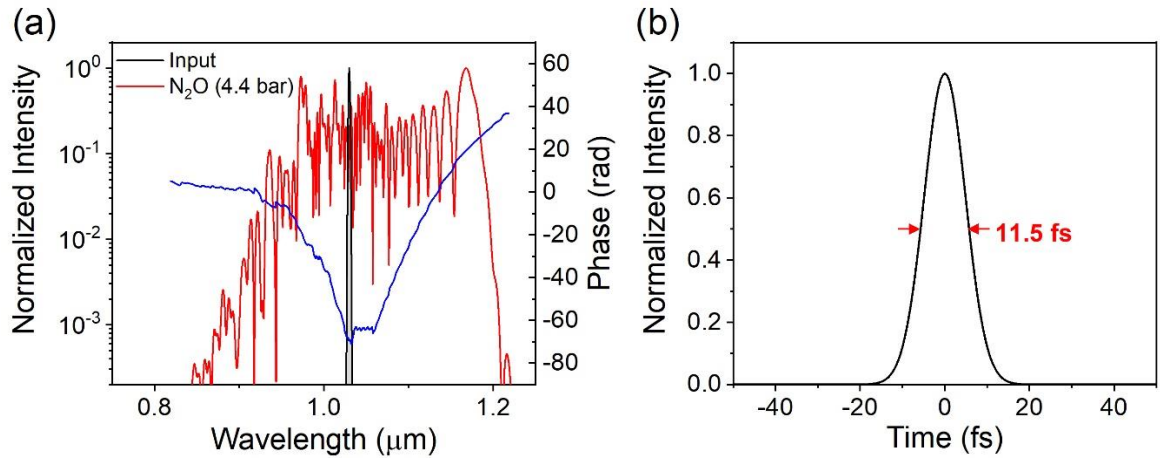

**Fig. S9.** Simulated propagation of 800 fs 1030 nm pulses in 4.4 bar of $N_2O$. (a) Normalized calculated spectrum after propagation in $N_2O$ with dispersion of $N_2$ (red line) and resultant spectral phase (blue line). The black line shows the input laser spectrum. (b) Bandwidth-limited intensity profile, showing a FHWM duration of 11.5 fs.

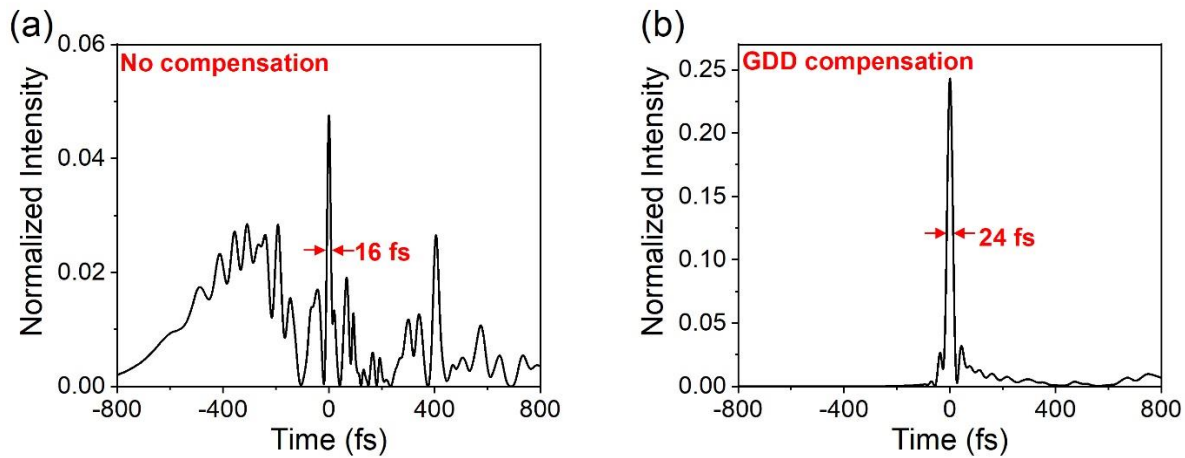

**Fig. S10.** Simulated intensity profiles resulting from propagation of 800 fs 1030 nm pulses in 4.4 bar of $N_2O$. (a) Uncompensated intensity profile showing a self-compressed peak with a FWHM duration of 16 fs, but with significant satellite pulse content. (b) Intensity profile obtained by adding only constant group delay dispersion, yielding a relatively clean pulse with FWHM duration of 24 fs. In both plots, the peak intensity is normalized to that of the transform-limited pulse.

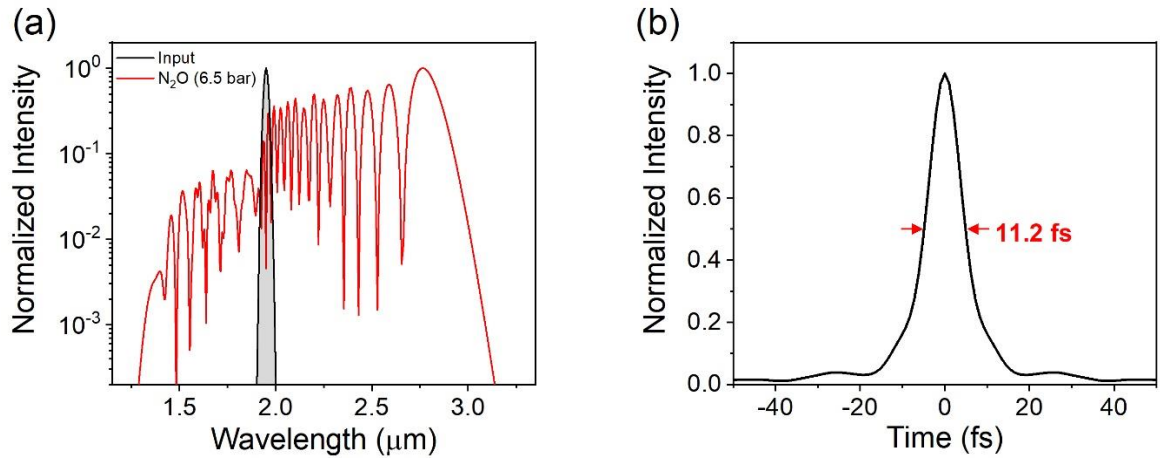

**Fig. S11.** Simulated propagation of 200 fs 1.95 μm pulses in 6.5 bar of $N_2O$. (a) Normalized calculated spectra after propagation in $N_2O$ with dispersion of $N_2$ (red line). The spectral weight is more heavily weighted towards longer wavelengths, and more than 92% of the energy is contained in the red-shifted region with wavelengths longer than 2.0 μm. The black line shows the input laser spectrum. (b) Bandwidth-limited intensity profile, showing a FHWM duration of 11.2 fs.

|  | $N_2$ (CMs) | $N_2O$ (CMs) | $N_2O$ (AOPDF) |
|---|---|---|---|
| **Input Pulse Energy** | 400 µJ | 140 µJ | 400 µJ |
| **Repetition Rate** | 250 Hz | 250 Hz | 1 kHz |
| **Fiber Output Pulse Energy** | 250 µJ (63%) | 84 µJ (60%) | 104 µJ (26%) |
| **Gas Pressure** | 6.5 bar | 2.4 bar | 3.0 bar |

**Table S1.** Laser parameters used for pulse compression results in Figs. S3-S5.